\documentclass[12pt]{article}

\usepackage{amssymb,amsmath,amsfonts,amsbsy,epsfig,psfrag}

\begin{document}

\begin{titlepage}

\begin{center}

\vspace*{-10ex}
\hspace*{\fill} KAIST-TH/2005-16

\vskip 1.5cm

\Huge{Dark energy in hybrid inflation}

\vskip 1cm

\large{ Jinn-Ouk Gong$^{1,2}$ \hspace{0.2cm} Seongcheol
Kim$^1$\footnote{sckim@muon.kaist.ac.kr}
\\ \vspace{0.5cm} {\em ${}^1$ Department of Physics, KAIST, Daejeon, Republic of
Korea
\\ \vspace{0.2cm}
${}^2$ International Center for Astrophysics, KASI, Daejeon, Republic of
Korea\footnote{Present address}} }

\vskip 0.5cm

\today

\vskip 1.2cm

\end{center}

\begin{abstract}

The situation that a scalar field provides the source of the accelerated expansion
of the universe while rolling down its potential is common in both the simple models
of the primordial inflation and the quintessence-based dark energy models. Motivated
by this point, we address the possibility of causing the current acceleration via
the primordial inflation using a simple model based on hybrid inflation. We trigger
the onset of the motion of the quintessence field via the waterfall field, and find
that the fate of the universe depends on the true vacuum energy determined by
choosing the parameters. We also briefly discuss the variation of the equation of
state and the possible implementation of our scenario in supersymmetric theories.

\end{abstract}

\end{titlepage}

\setcounter{page}{0}
\newpage
\setcounter{page}{1}

\section{Introduction}
\label{intro}

The discovery of the acceleration of the universe by the measurements of the
luminosity-redshift relation for type Ia supervonae \cite{sn}, combined with the
observations of the anisotropies in the cosmic microwave background (CMB)
\cite{cmb}, confirmed that about 70\% of the density of the universe is made up of
an unknown form of energy. This component, usually referred to as ``dark energy", is
unclumped and smoothly distributed, with an exotic property that it has negative
pressure to cause the expansion of the universe to accelerate. The simplest
candidate of dark energy is a non-vanishing, positive cosmological constant
\cite{ccp}. We have to, however, explain why it is non-zero but vanishingly small
($\sim 10^{-12} \mathrm{eV}^4$): possibly we can resort to some yet unknown
fundamental symmetry which makes the cosmological constant vanishingly small but
non-zero, or we might invoke an anthropic consideration that the observed small
value of the cosmological constant allows life and that is why we observe
it\footnote{Interestingly, this idea is easily fulfilled in modern string theory
landscape \cite{landscape}: in string theory, there exist many (more than
$10^{1500}$ \cite{stringsol}) different vacua, and some of them may be suitable for
some kind of intelligent observers like us.}.

An alternative form of dark energy is a slowly rolling scalar field, called
quintessence, which has not yet relaxed at its ground state \cite{Q}. For recent
years, there have been considerable developments in the dynamics of the quintessence
field. For example, in the context of so-called tracker solutions \cite{tracker},
the simplest case arises for a potential of the form $V = M^{4 + n} \phi^{-n}$,
where $n$ is positive and $M$ is an adjustable constant. By choosing $M$ suitably,
it is possible to make a transition from an early matter-dominated universe to a
later quintessence-dominated universe, free from fine tuning of the initial
conditions. Another interesting possibility, $k$-essence, is to introduce a
non-canonical kinetic term for the scalar field \cite{kessence}, which makes the
evolution of the scalar field dependent on the background equation of state,
explaining why dark energy is dominating now. Anyway, apart from the details, the
situation that a scalar field is slowly rolling down its potential is reminiscent of
the primordial inflation \cite{inf}, where a scalar field (the inflaton) provides
the vacuum-like energy density ($p \approx -\rho$) necessary for a phase of the
accelerated expansion by slowly rolling down its flat potential.

Hence one may naturally raise a question of how we can couple the early accelerated
expansion, the primordial inflation, and the current one together \cite{qinflation}:
is it possible to cause the acceleration we observe recently by the primordial
inflation? In this paper, we are going to discuss this possibility using a simple
model based on the hybrid inflation \cite{hybrid}, which arises naturally in many
string-inspired inflation models, in particular in potentials for moduli fields.
This paper is organized as follow: in Section~\ref{model}, we present a simple model
of dark energy and analyze its dynamics in detail, presenting several conditions
which should be satisfied for our model to work properly. Unlike the conventional
lore that the true minimum of the quintessence potential is presumed to vanish, we
find that positive, zero, or negative vacuum energy is possible as we choose a
different set of parameters. In Section~\ref{discussions}, we discuss the variation
of the equation of state $w$ which might be detected in future observations such as
HETDEX \cite{hetdex}. Also we briefly address the possibility of realizing our model
in supersymmetric theories and obstacles to overcome. We briefly summarize this
paper in Section~\ref{summary}.

\section{A model of dark energy}
\label{model}

In this section, we discuss a simple model of dark energy based on the hybrid model
of inflation \cite{hybrid}. Perhaps the simplest way to combine the onset of present
acceleration of the universe with the primordial inflation is to directly couple the
inflaton with the quintessence field. In this case, however, first we should ensure
that inflation lasts for a long enough time (at least 60 $e$-foldings) without being
disturbed by the interaction with the quintessence field. Moreover, the quintessence
potential must remain extremely flat after the inflaton reaches its minimum and
decays to reheat the universe. Rather, it is more plausible to couple the
quintessence field with some different field which plays no role during the inflaton
rolls down its potential and only works to finish inflation: this is what the
waterfall field in hybrid inflation model does. Hence we can write the effective
potential as
\begin{equation}\label{potential}
V(\phi,\psi,\sigma) = \frac{1}{2}m^2\phi^2 + \frac{g^2}{2}\phi^2\psi^2 -
\frac{h^2}{2}\psi^2\sigma^2 + \frac{1}{4\lambda}(M_\psi^2 - \lambda\psi^2)^2 +
\frac{1}{4\mu}(M_\sigma^2 + \mu\sigma^2)^2 \, ,
\end{equation}
where we take $\phi$ as the inflaton field, $\psi$ as the waterfall field, and
$\sigma$ as the quintessence field. As can be seen from the coupled terms above, we
will experience two phase transitions, which distinguish different stages of the
evolution of the universe. In the following subsections, we will discuss each stage
in detail. 

\subsection{First stage: primordial inflation}
\label{1st}

The effective masses squared of $\psi$ and $\sigma$ are $g^2\phi^2 - M_\psi^2$ and
$-h^2\psi^2 + M_\sigma^2$, respectively. Hence, for $\psi < \psi_c \equiv
M_\sigma/h$, the only minimum of the effective potential, Eq.~(\ref{potential}), is
at $\sigma = 0$. Also, with $\sigma = 0$, for $\phi > \phi_c \equiv M_\psi/g$, the
only minimum of the effective potential is at $\psi = 0$. Thus, at the early stage
of the evolution of the universe, $\psi$ and $\sigma$ are trapped at 0 while $\phi$
remains much larger than $\phi_c$ for a long time. This stage lasts until $\phi =
\phi_c$, and at that moment we assume that the vacuum energy density $V(0,0,0) =
M_\psi^4/(4\lambda) + M_\sigma^4/(4\mu)$ is much larger than the potential energy
density of the inflaton field $m^2\phi_c^2/2 = m^2M_\psi^2/(2g^2)$, so that the
Hubble parameter is given by
\begin{equation}\label{H1stlast}
H^2 \simeq \frac{1}{12m_\mathrm{Pl}^2} \left( \frac{M_\psi^4}{\lambda} +
\frac{M_\sigma^4}{\mu} \right) \, ,
\end{equation}
where $m_\mathrm{Pl}^2 = (8\pi G)^{-1}$ is the reduced Planck mass. We can
additionally assume that the vacuum energy $V(0,0,0)$ is dominated by
$M_\psi^4/\lambda$, i.e.,
\begin{equation}\label{psidomination}
\frac{M_\psi^4}{\lambda} \gg \frac{M_\sigma^4}{\mu} \, ,
\end{equation}
so that the subsequent evolutions of the universe after inflation, e.g., reheating,
becomes identical to the usual hybrid model. This immediately gives
\begin{equation}\label{mbound1}
m \ll \frac{g}{\sqrt{2\lambda}}M_\psi \, .
\end{equation}
Now, combining Eqs.~(\ref{H1stlast}) and (\ref{psidomination}), the slow-roll
condition for $m$ gives
\begin{equation}\label{Mpsibound1}
M_\psi^2 \gg 6\sqrt{\lambda}mm_\mathrm{Pl} \, .
\end{equation}
Note that this gives another bound on $m$ as $m \ll
M_\psi^2/(6\sqrt{\lambda}m_\mathrm{Pl})$, and the tightest bound on $m$ depends on
the comparison between the prefactors of this expression and Eq.~(\ref{mbound1}),
i.e., $g$ and $M_\psi/(3\sqrt{2}m_\mathrm{Pl})$.

\subsection{Second stage: phase transition between $\phi$ and $\psi$}
\label{2nd}

As discussed above, when $\phi$ becomes smaller than $\phi_c$, a phase transition
with the symmetry breaking for $\psi$ occurs. To finish inflation as soon as $\phi$
reaches the critical value $\phi_c$, i.e., for the so-called ``waterfall", we first
require that the absolute value of the effective mass squared of $\psi$ be much
larger than $H^2$. During inflation, using the slow-roll equation of motion
$3H\dot\phi + m^2\phi = 0$ and Eqs.~(\ref{H1stlast}) and (\ref{psidomination}), we
can find that when the inflaton $\phi$ reaches $\phi_c$, it decreases by
\begin{equation}
\Delta\phi = \frac{4\lambda m^2m_\mathrm{Pl}^2}{gM_\psi^3}
\end{equation}
during the time interval $H^{-1}$. At that time, the absolute value of the effective
mass squared of $\psi$ is given by
\begin{equation}\label{psimass}
|m_\psi^2| = \frac{8\lambda m^2m_\mathrm{Pl}^2}{M_\psi^2} \, ,
\end{equation}
which is much greater than $H^2$ for
\begin{equation}\label{waterfall1}
M_\psi^3 \ll 12\sqrt{2}\lambda mm_\mathrm{Pl}^2 \, .
\end{equation}
Also, we demand that the time scale for $\phi$ to roll down from
$\phi_c$ to 0 be much shorter than $H^{-1}$: let us denote $\Delta
t$ as the time $\phi$ takes as it moves from $\phi_c$ to 0. Then,
from
\begin{align}
\frac{\partial V}{\partial\phi} & = m^2\phi + g^2\phi\psi^2
\nonumber \\
& \simeq \frac{8g^2m^2m_\mathrm{Pl}^2\phi}{M_\psi^2} \, ,
\end{align}
where we have used Eq.~(\ref{psimass}) and the slow-roll equation of $\phi$, we have
\begin{equation}
\left| \frac{H}{\Delta t} \right| = \frac{8g^2m^2m_\mathrm{Pl}^2}{3M_\psi^2} \gg H^2
\, ,
\end{equation}
which gives another condition
\begin{equation}\label{waterfall2}
M_\psi^3 \ll 4\sqrt{2\lambda}gmm_\mathrm{Pl}^2 \, .
\end{equation}
Note that the parameters with respect to $\phi$-$\psi$ transition
are further constrained from the observed COBE amplitude of
density perturbations: we obtain \cite{hybrid}
\begin{equation}\label{pertconst}
M_\psi^3 \ll 5 \times 10^{-5} \lambda g^{-1}mm_\mathrm{Pl} \, ,
\end{equation}
which looks similar to Eqs.~(\ref{waterfall1}) and (\ref{waterfall2}). Combining
Eqs.~(\ref{waterfall1}), (\ref{waterfall2}) and (\ref{pertconst}), we can extract
the valid range of the parameters $m$, $M_\psi$, $g$ and $\lambda$: e.g. if we take
$g^2 \sim \lambda \sim 10^{-1}$, $m \sim 10^2 \mathrm{GeV}$ and $M_\psi \sim 10^{11}
\mathrm{GeV}$, all the conditions are satisfied. After this phase transition, the
waterfall field $\psi$ oscillates at the minimum and decays so that the universe is
reheated\footnote{At this point, although we require that small enough be the
coupling $h^2$ of $\sigma$ to $\psi$, and in turn to the thermal bath, the
oscillation of $\psi$ may affect the dynamics of $\sigma$. However, we note that
because of the negative sign of their interaction term in Eq.~(\ref{potential}), the
pattern of instability is quite different from the usual parametric resonance
\cite{parametric} even though the interaction is strong enough: we will study this
issue in more detail separately.}.

With these conditions, as soon as $\phi$ reaches $\phi_c$, inflation ends within a
Hubble time, $H^{-1}$. Also note that when such a rapid phase transition occurs
within observationally interesting range, still we can calculate the power spectrum
and the spectral index for the density perturbations \cite{gsr} by generalizing the
standard perturbative method \cite{gs}: generally, the power spectrum shows a scale
dependent oscillations after the phase transition \cite{feature}.

\subsection{Third stage: phase transition between $\psi$ and $\sigma$}
\label{3rd}

While $\psi$ is rolling down to $\psi_0 \equiv M_\psi/\sqrt{\lambda}$, the effective
mass squared of $\sigma$ might become negative then another phase transition occurs.
For this to happen, we want $\psi_c < \psi_0$, unless no instability for $\sigma$
would develop. This is equivalent to the condition\footnote{It should be guaranteed
that the effective potential is bounded from below so that $\psi$ is settled at
$\psi_0$. When $\psi$ moves along $\psi = \sigma$ direction, this is equivalent to
the condition $\lambda
> 2h^2$. This means, combined with Eq.~(\ref{psisigma}), $M_\psi$
is (much) bigger than $M_\sigma$. We are grateful to Andrei Linde
for pointing out this.}
\begin{equation}\label{psisigma}
\frac{M_\sigma}{h} < \frac{M_\psi}{\sqrt{\lambda}} \, .
\end{equation}
Note that we can obtain the same condition by requiring that the minimum along
$\sigma$ direction, $\sigma_0$ given by
\begin{equation}
\left. \frac{\partial V}{\partial\sigma} \right|_{\sigma_0} =
-h^2\psi^2\sigma_0 + M_\sigma^2\sigma_0 + \mu\sigma_0^3 = 0 \, ,
\end{equation}
be real, i.e.,
\begin{equation}
\sigma_0^2 = \frac{h^2M_\psi^2 - \lambda M_\sigma^2}{\lambda\mu} > 0 \, .
\end{equation}
Once the above condition is satisfied and $\sigma$ rolls down the effective
potential, after all the fields are settled at
\begin{equation}
\phi = 0 \, , \hspace{0.5cm} \psi = \psi_0 \, , \hspace{0.5cm} \mbox{and}
\hspace{0.5cm} \sigma = \sigma_0 \, ,
\end{equation}
where the potential becomes
\begin{equation}\label{globalmin}
V_0 = \frac{h^2M_\psi^2}{4\lambda^2\mu} \left(2\lambda M_\sigma^2 - h^2M_\psi^2
\right) \, .
\end{equation}

\subsubsection{Positive vacuum density}

When $V_0 > 0$, Eq.~(\ref{globalmin}) is greater than zero and it corresponds to a
non-zero, positive vacuum energy. Hence no matter $\sigma$ evolves quickly or not,
we are provided with the source of the current acceleration of the universe in this
case, and this acceleration will last forever: the universe will eventually behave
as a de Sitter space. Note that the possible maximum value of $V_0$,
$h^2M_\psi^2M_\sigma^2/(4\lambda\mu)$, could be at most as large as the current
critical density, $\rho_\mathrm{crit} \sim \left( 10^{-12}\mathrm{GeV} \right)^4$.
However, an extreme fine tuning is required to match this value: if we restrict our
interest to this case only, our model is no better than the conventional
$\Lambda$CDM because the latter is simpler and hence preferable. Anyway here we do
not try to improve this situation. A study of alleviating this fine tuning problem
is very important and interesting, but is outside the scope of the present paper.

\subsubsection{Zero vacuum density}

For the case
\begin{equation}\label{V0=0condition}
2\lambda M_\sigma^2 = h^2M_\psi^2 \, ,
\end{equation}
Eq.~(\ref{globalmin}) is exactly zero and no vacuum energy exists. This case
corresponds to the usual quintessential inflation \cite{qinflation}, where the
cosmological constant $\Lambda$ is assumed to be zero due to some unknown symmetry,
and the observed dark energy is supplied by e.g. another scalar field, here
$\sigma$. To be able to explain the acceleration of the universe, we require that
still $\sigma$ be rolling down so that the potential is non-zero. That is, $\sigma$
should roll extremely slowly. Note that although $\sigma$ begins to evolve only
after $\psi$ reaches $\psi_c$, for the most time during $\sigma$ rolls down the
effective potential to $\sigma_0$, $\psi$ has already settled at $\psi_0$ within a
time $\Delta t \ll H^{-1}$ as soon as $\phi \sim \phi_c$. Thus, we just set $\psi =
\psi_0$ throughout the evolution of $\sigma$.

Under this circumstance, when $\sigma$ begins to roll, $H$ is given by
\begin{equation}\label{sigmarollH}
H^2 \simeq \frac{M_\sigma^4}{12\mu m_\mathrm{Pl}^2} \, .
\end{equation}
For $\sigma$ to slowly evolve, we require that $|m_\sigma^2|$, the absolute value of
the effective mass squared of $\sigma$, be much smaller than $H^2$, from which we
obtain a condition
\begin{equation}\label{V0=0con1}
M_\sigma \gg 6\sqrt{\mu}m_\mathrm{Pl} \, ,
\end{equation}
where we have used Eq.~(\ref{V0=0condition}). Indeed, the same
bound could be found when we also require that $\sigma$ roll very
slowly so that for $\sigma$ to move by an infinitesimal
displacement $\Delta\sigma$, it takes much longer time than
$H^{-1}$: the effective potential is given by
\begin{align}\label{potential_sigma}
V \left(\phi = 0, \psi = \psi_0, \sigma \right) & = \frac{M_\sigma^4}{4\mu} +
\frac{1}{2} \left( M_\sigma^2 - \frac{h^2}{\lambda}M_\psi^2 \right) \sigma^2 +
\frac{1}{4}\mu\sigma^4
\nonumber \\
& = \frac{\mu}{4} \left( \sigma^2 - \frac{M_\sigma^2}{\mu} \right)^2 \, ,
\end{align}
where we have used Eq.~(\ref{V0=0condition}). Hence, combining with the
slow-roll equation
\begin{equation}\label{srequation}
3H\dot\sigma + \frac{\partial V}{\partial\sigma} \simeq 0 \, ,
\end{equation}
we find
\begin{equation}
\frac{\dot{\sigma}}{\sigma} \simeq \frac{M_\sigma^2}{3H} \, ,
\end{equation}
i.e. $\delta\sigma/\sigma \simeq M_\sigma^2/(3H^2)$ for $\delta t \simeq H^{-1}$.
Requiring $\delta\sigma/\sigma \ll 1$ gives
\begin{equation}\label{V0=0con2}
M_\sigma \gg 6\sqrt{\mu}m_\mathrm{Pl} \, ,
\end{equation}
which is the same as Eq.~(\ref{V0=0con1}).

\subsubsection{Negative vacuum density}

Recent observations find that the vacuum energy, or cosmological constant, is very
small and positive, with its value being of order $10^{-12} \mathrm{eV}^4$. The case
of negative cosmological constant is not observationally justified, which seems to
support the claim that the true minimum of the quintessence potential is zero. The
vacuum state with negative energy is, nevertheless, an interesting theoretical
possibility, especially in string theories where anti de Sitter solutions are
popular. Hence a negative potential may play an important role in cosmology
motivated from string theory or M theory, such as cyclic universe model
\cite{cyclic}.

We obtain $V_0 < 0$ when $V(\phi = 0, \psi = \psi_0, \sigma =
\sigma_t) = 0$ has a real solution and that this solution is
smaller than $\sigma_0$. If we take the limit $h^2M_\psi^2/\lambda
\gg M_\sigma^2$, the effective mass squared of $\sigma$ is given
by
\begin{equation}
m_\sigma^2 \simeq -\frac{h^2M_\psi^2}{\lambda} \, .
\end{equation}
Imposing the slow-roll condition, this gives
\begin{equation}
M_\sigma \gg 6\sqrt{\mu}m_\mathrm{Pl} \, ,
\end{equation}
the same as Eqs.~(\ref{V0=0con1}) and (\ref{V0=0con2}).

Since the dark energy observed now is positive definite, $\sigma$ should be still
evolving and not have crossed $\sigma_t$: thus we again require that $\sigma$ roll
down very slowly, from which we obtain the same conditions as the previous section,
Eqs.~(\ref{V0=0con1}) and (\ref{V0=0con2}). However, unlike the cases before, the
true vacuum energy is negative and after a (tremendously) long time the universe
will collapse eventually \cite{negativeV}, no matter how small the magnitude is.

\section{Discussions}
\label{discussions}

\subsection{The equation of state}
\label{w}

Generally, for evolving dark energy models the equation of state $w$ is not a
constant but a slowly evolving function. To provide an acceleration of the universe
it is constrained to be smaller than $-1/3$, usually taken to be $w < -2/3$. Recent
combined observations of WMAP and SDSS constrains $w$ to be $-1$, with uncertainties
at the 20\% level \cite{wmapsdss}. For the first case discussed in the previous
section where $V_0 > 0$, this is easily satisfied: $w$ is very slowly varying as
$\sigma$ rolls down its potential, finally becoming exactly $-1$ at the absolute
minimum $V_0$. When $V_0 = 0$, $w$ will be eventually 0 after $\sigma$ settles at
$\sigma_0$. For the third case, however, it is not as trivial as the other cases.
Naively, we may take $\dot\sigma$ to be non-zero although very small, then we can
write
\begin{equation}\label{wV/K}
w = \frac{1 - V/K}{1 + V/K} \, ,
\end{equation}
where $K$ stands for the kinetic energy, $\dot\sigma^2/2$. At first look, it
seems that $w$ is divergent at $V = -K$ and a discontinuity appears and moreover
a phantom phase $w < -1$ exists after that point. Let us more closely see this.

\begin{figure}[h]
\begin{center}
\psfrag{w}{$w$}%
\psfrag{t}{$t$}%
\psfrag{z}{$z$}%
\epsfig{file=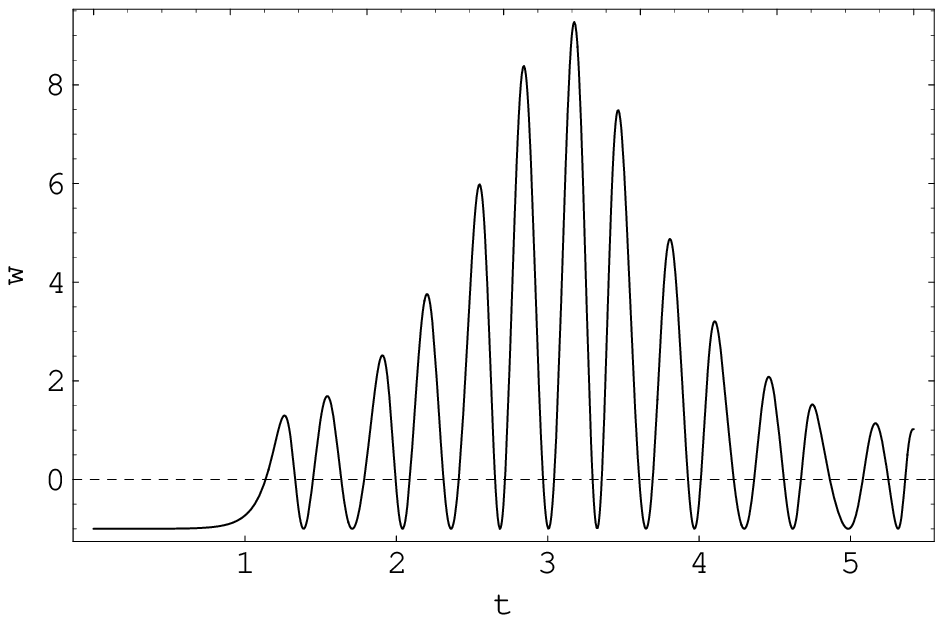, width = 8.1cm}%
\epsfig{file=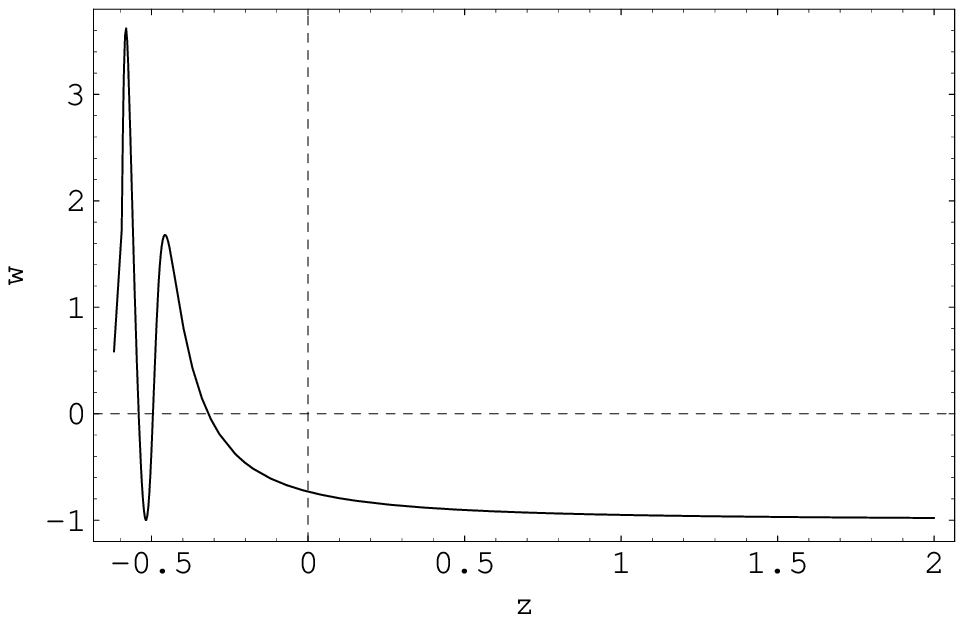, width = 8.1cm}%
\caption{\label{fig_w}(Left) plot of $w$ versus time in the unit
of $t_0$ which corresponds to $z = 0$, and (right) $w$ versus
redshift $z$. As $\sigma$ approaches $\sigma_t$ where $V = 0$, $w$
increases and becomes positive. When $\sigma$ reaches the minimum
of its potential which is negative, $\sigma$ oscillates around
this minimum and $w$ shows the corresponding oscillatory behavior.
When the universe expands no more and begins to contract, the
amplitude of this oscillation decreases.}
\end{center}
\end{figure}

The simplest way is to solve the relevant equations numerically: we choose
\begin{align}
\ddot\sigma + 3 \frac{\dot a}{a}\dot\sigma + \frac{\partial V}{\partial\sigma} &
= 0 \, ,
\nonumber \\
\frac{\ddot a}{a} & = \frac{1}{3m_\mathrm{Pl}^2} (V - \dot\sigma^2) \, ,
\end{align}
to solve for $\sigma(t)$ and $a(t)$. In Figure~\ref{fig_w}, we plot the resulting
equation of state $w$ where, as is shown, it oscillate. To understand this behavior
first let us consider the expansion of the universe with such a negative potential
\cite{negativeV}: the universe keeps expanding until the moment $\rho =
\dot\sigma^2/2 + V = 0$ where the scale factor reaches its largest value, and the
universe shrinks afterwards. From the Friedmann equation describing a flat universe
$H^2 = \rho/(3m_\mathrm{Pl}^2)$ which has no solution with $\rho < 0$\footnote{This
is the reason why we cannot reach an anti de Sitter universe dominated by a negative
cosmological constant \cite{fastroll}.}, we can see that to maintain a real value of
$H$ no matter positive or negative, $\rho$ should remain positive. Hence, when $V$
is negative, the kinetic energy must compensate this negative potential energy and
consequently these energies oscillate out of phase: for example, at the negative
bottom of the potential, the field is moving fastest so that we have maximum kinetic
energy here so that $w$ becomes very large. This is the reason why the equation of
state $w$ is oscillating\footnote{Note that this oscillation occurs only when the
potential is bounded from below: if this is not the case, e.g. when $V = V_0 -
m^2\phi^2/2$, we don't see such an oscillation.} and sometimes becomes greater than
1, the ``stiff'' fluid. We can easily find the relation with respect to the redshift
$z$, by making use of the relation $a \propto (1 + z)^{-1}$: by solving the
equations
\begin{align}
\frac{\dot a}{a} (1 + z) a' & = -\dot a \, ,
\nonumber \\
\frac{\dot a}{a} (1 + z) \left( \dot a \right)' & = \frac{a}{3m_\mathrm{Pl}^2}
\left( \dot\sigma^2 - V \right) \, ,
\nonumber \\
\frac{\dot a}{a} (1 + z) \sigma' & = -\dot\sigma \, ,
\nonumber \\
\frac{\dot a}{a} (1 + z) \left( \dot\sigma \right)' & = -3 \frac{\dot a}{a}
\dot\sigma - \frac{\partial V}{\partial\sigma} \, ,
\end{align}
where a prime denotes the derivative with respect to $z$, we can obtain $w$ in
terms of $z$. A typical relation between $w$ and $z$ is shown in the right panel
of Figure~\ref{fig_w}.

\subsection{Supersymmetric realization}
\label{susy}

Supersymmetry is believed to be the most promising candidate related to the
fundamental problems in particle physics, such as the hierarchy problem and the
gauge coupling unification. Since we have not yet observed any supersymmetric
partner of known particles, supersymmetry is broken. In the primordial universe
where the energy scale is much higher than the present one, however, the rich
structure of supersymmetric and supergravity theories should have played an
important role. Hence it is natural to try to implement inflationary scenario within
supersymmetry. There has been an encouraging progress on the hybrid inflation
scenario in the context of supersymmetric theories
\cite{susyhybrid,dterminf,hybridsugra}. For example, consider a simple
superpotential \cite{hybridsugra}
\begin{equation}
W = \phi \left( \lambda\psi_1\psi_2 - \mu^2 \right) \, ,
\end{equation}
where $\psi_1$ and $\psi_2$ are a pair of superfields in non-trivial
representations of some gauge group under which $\phi$ is neutral. Then, in a
globally supersymmetric theory, the effective potential is given by
\begin{equation}
V = \lambda^2 |\phi|^2 \left( |\psi_1|^2 + |\psi_2|^2 \right) + \left|
\lambda\psi_1\psi_2 - \mu^2 \right|^2 \, . 
\end{equation}
Here, the absolute supersymmetric minimum appears at $\phi = 0, \psi_1 = \psi_2 =
\mu/\sqrt{\lambda}$. However, for $\phi > \phi_c = \mu/\sqrt{\lambda}$, $\psi_1$ and
$\psi_2$ obtain positive masses squared and hence are confined at the origin. By
simply adding a mass term $m^2\phi^2/2$ which softly breaks supersymmetry, we see
that the hybrid inflation scenario is possible. If we take into account radiative
corrections instead, the total effective potential including one-loop corrections is
\begin{equation}
V = \mu^4 \left[ 1 + \frac{\lambda^2}{8\pi^2} \ln \left( \frac{\phi}{\phi_c} \right)
+ \cdots \right]
\end{equation}
when $\phi \gg \phi_c$, and we can see that still inflation is possible.

Similarly, we can hope to implement our scenario within supersymmetry. The simplest
possibility should be $D$-terms since, as we can see from Eq.~(\ref{potential}), the
coupled term of $\psi$ and $\sigma$ has a negative sign, so that an instability for
$\sigma$ could be developed as $\psi$, initially confined at zero, rolls away from
the origin. This is easily achieved by assuming that $\psi$ and $\sigma$ are
oppositely charged under some gauge symmetry. If it is a $U(1)$ symmetry, we can
write the $D$-term contribution as
\begin{equation}\label{dterm1}
\frac{\alpha^2}{2} \left( - |\psi|^2 + |\sigma|^2 + \xi_1 \right)^2 \, ,
\end{equation}
where we assumed that $\psi$ and $\sigma$ have charges equal to $-1$ and $+1$
respectively, $\alpha$ is the gauge coupling, and $\xi_1$ is a Fayet-Iliopoulos
$D$-term. We may also include another gauge symmetry under which $\phi$ and
$\psi$ are charged with the same sign, then $D$-term is given by
\begin{equation}\label{dterm2}
\frac{\beta^2}{2} \left( |\phi|^2 + |\psi|^2 + \xi_2 \right)^2 \, ,
\end{equation}
where we have introduced another gauge coupling $\beta$ and Fayet-Iliopoulos
term $\xi_2$. Taking only Eqs.~(\ref{dterm1}) and (\ref{dterm2}) into account,
the coupled terms are given by
\begin{equation}
\left[ \beta^2|\phi|^2 - \left( \alpha^2\xi_1 - \beta^2\xi_2 \right) \right]
|\psi|^2 + \left( -\alpha^2|\psi|^2 + \alpha^2\xi_1 \right) |\sigma|^2 \, ,
\end{equation}
we can reproduce the same forms for $\psi^2$ and $\sigma^2$ terms as
Eq.~(\ref{potential}) as long as $\alpha^2\xi_1 - \beta^2\xi_2$ is positive.
This sheds some light on the realization of Eq.~(\ref{potential}).

However, this is far from a realistic possibility. First of all, we have not
considered any $F$-term contributions. We do not want to couple $\phi$ and $\sigma$
to guarantee their desired behaviors on the flat enough effective potential, but it
seems not easy to achieve this through the contributions from $F$-term. More
fundamentally, in globally supersymmetric theories the scalar potential is either
positive or zero, which does not include the interesting case of $V_0 < 0$. Also, it
is thought that below the Planck scale particle physics is described by an effective
$N = 1$ supergravity theory derived from string theory. Hence, a more detailed
analysis of our model and associated problems should be addressed in the context of
supergravity: for example, the mass of the field should be very small, of order the
present Hubble parameter $H_0 \sim 10^{-42} \mathrm{GeV}$, so that it is still
rolling toward its true minimum. However, usually scalar fields acquire masses of
order the gravitino mass $m_{3/2}^2$, much heavier than $H_0$. One way to evade this
difficulty is to use pseudo Nambu-Goldstone bosons\footnote{Note that this also
naturally solves the $\eta$ problem associated with generic scalar field potentials
in supergravity.} \cite{pngb}, e.g., string/M theory axion, as $\sigma$ field
\cite{axion}. We will leave such an analysis as a challenging future work.

\section{Summary}
\label{summary}

We have investigated a simple dark energy model based on hybrid inflation. The
quintessence field $\sigma$ is coupled to the waterfall field $\psi$ so that as
$\psi$ rolls towards $\psi_0 = M_\psi/\sqrt{\lambda}$, $\sigma$ begins to move along
the effective potential provided that Eq.~(\ref{psisigma}) is satisfied. A number of
bounds on the parameters of the model are found, such as Eqs.~(\ref{psidomination}),
(\ref{mbound1}), (\ref{Mpsibound1}), (\ref{waterfall1}) and (\ref{waterfall2}). An
interesting point is that the true minimum of the effective potential,
Eq.~(\ref{globalmin}), depends on our choice of the parameters, allowing the vacuum
state with positive, negative and zero energy. When it is negative, the universe
will eventually collapse, showing an oscillation in the equation of state $w$ for
dark energy which we can hope to detect in future observations. This model could be
realized in supersymmetric theories via $D$-term contributions, but including
$F$-term parts and supergravity effects makes this model not easy to be achieved.

\subsection*{Acknowledgements}
We thank Kiwoon Choi, Salman Habib, Andrei Linde and Carlos Mu\~noz for helpful
discussions and suggestions. We are also indebted to the anonymous referee for many
important and invaluable comments. JG is grateful to the SF05 Cosmology Summer
Workshop, and SK to the Summer Institute 2005 for hospitalities when this work was
in progress. This work was supported in part by the Astrophysical Research Center
for the Structure and Evolution of the Cosmos funded by the Korea Science and
Engineering Foundation and the Korean Ministry of Science, the Korea Research
Foundation grant KRF PBRG 2002-070-C00022, and the Brain Korea 21.

\end{document}